\documentclass[aps,prl,reprint,floatfix,nofootinbib,superscriptaddress,longbibliography]{revtex4-2}
\usepackage{amsmath,amssymb}
\usepackage{bm}
\usepackage{graphicx}
\usepackage[breaklinks=true,colorlinks=true,citecolor=blue,linkcolor=blue,urlcolor=blue]{hyperref}
\DeclareMathOperator{\arcosh}{arcosh}
\begin{document}

\title{Analytic Origin of Green-Function Compression in the Intermediate Representation}

\author{Takahiro Misawa}
\affiliation{Institute for Solid State Physics, University of Tokyo}

\date{\today}

\begin{abstract}
Information compression plays a central role in diverse fields of
modern science and technology, from communication theory to machine
learning. In condensed-matter physics, the intermediate
representation (IR) basis has recently been developed as an
efficient method for compressing imaginary-time Green functions,
which are fundamental quantities for describing quantum many-body
systems.\ This compression relies on the rapid decay of the singular values with
the basis index and the unusually weak growth of the effective rank
with inverse temperature.\ Because of these useful features, the
IR basis is now widely used as a standard method in quantum many-body
calculations.\ However, the analytic origin of its compression capability has remained unclear.
Here we uncover a finite-Laplace-transform structure underlying
the IR kernel, which reveals that the eigenfunctions of the IR kernel
admit a natural expansion in terms of classical special functions, the
oblate spheroidal wave functions.\ This finite-Laplace-transform structure also enables us to analytically clarify the compression
mechanism of the IR basis.\ Our results provide a mathematical foundation for the compression of
imaginary-time Green functions, connecting quantum many-body physics
with theories of information compression and finite integral transforms.
\end{abstract}

\maketitle

\textit{Introduction.}---Efficient representation of data is a common
foundation of modern science and technology.  It reduces the cost of
storing, transmitting, and analyzing information across fields
ranging from communication theory to machine learning and scientific
computing~\cite{Shannon1948,Carleo2019,Schollwock2011}.  Efficient
compression becomes possible when data have a hidden mathematical
structure that restricts the relevant degrees of freedom to a small
subspace.  Familiar examples include image compression exploiting the
rapid decay of wavelet coefficients for natural images~\cite{Mallat1989}.
In a more analytic setting, a classical example is the
time-bandwidth concentration problem studied by Slepian, Pollak, and
Landau~\cite{Slepian1961,LandauPollak1961,Slepian1983}.  It arose in
communication theory and later found applications in optics and
quantum information~\cite{Slepian1965Apodization,ImaiHayashi2009}.
This problem is governed by a finite Fourier transform, and the
analytic solvability of the resulting integral equation revealed how
time and band limitation produce an effectively finite-dimensional
subspace.

In condensed-matter physics, the intermediate representation (IR)
basis was introduced as an efficient method for compressing
imaginary-time Green
functions~\cite{Shinaoka2017,Otsuki2017,Chikano2018,Shinaoka2022}.
The IR basis is obtained from the singular value decomposition (SVD) of
the imaginary-time--frequency kernel.  Its usefulness originates
from the rapid decay of the singular values with the basis index.
Moreover, in the fermionic case, the effective rank grows only
logarithmically with the inverse temperature, whereas in the
bosonic case it saturates at low temperatures.  This compactness
has made the IR basis a standard computational method.  It has been widely
applied in quantum many-body calculations, including
sparse sampling for diagrammatic calculations~\cite{Li2020},
Migdal--Eliashberg calculations~\cite{Wang2020,Mori2024},
two-particle self-consistent calculations~\cite{GauvinNdiaye2024_TPSC},
fluctuation-exchange calculations~\cite{Witt2021_FLEX}, and
dynamical mean-field theory~\cite{Nagai2019}.  Numerical tests have been
performed extensively, and convenient numerical libraries such as
\texttt{irbasis}~\cite{Chikano2019} and
\texttt{sparse-ir}~\cite{Wallerberger2023} are now available.
However, what has been missing is an analytic mechanism, based on the
structure of the imaginary-time kernel, that explains why the effective
rank grows only logarithmically for fermions and saturates for bosons.
Existing analytic understanding has been limited to the
infinite-temperature limit, where the basis reduces to Legendre
polynomials~\cite{Shinaoka2017}.

In this Letter, we clarify the analytic structure of this problem by
reducing the IR singular-value problem to a generalized eigenproblem
generated by a finite Laplace transform.  We show that the resulting integral
operator possesses a commuting second-order differential operator,
whose eigenfunctions are the oblate spheroidal wave
functions~\cite{Flammer1957}.  This result gives an explicit
construction of the IR singular functions as infinite series in
the oblate spheroidal wave functions.  More importantly for compression,
this finite-Laplace-transform structure exposes the low-temperature mechanism
controlling the effective rank.  In the low-temperature limit, the
fermionic problem becomes a Hilbert-kernel problem on a logarithmic energy
interval, which explains the rapid decay of the singular values and the
logarithmic growth of the effective rank with inverse temperature.  Since the
bosonic frequency factor removes this logarithmic energy interval, the
effective rank saturates at low temperatures.  These results clarify the analytic origin of
IR compression of imaginary-time Green functions.

\textit{Setup.}---The imaginary-time Green function and the spectral
density are related at finite temperature by
\begin{equation}
G^\alpha(\tau)
=-\!\!\int_{-\omega_{\max}}^{\omega_{\max}}\!\!
K^\alpha(\tau,\omega)\rho^\alpha(\omega)\,d\omega .
\label{eq:G-stat}
\end{equation}
Here $\tau$ is the imaginary time with $0\le\tau\le\beta$
($\beta$ is the inverse temperature),
$\rho^\alpha(\omega)$ is the spectral density, $\alpha$ denotes
the particle statistics ($F$ for fermions and $B$ for bosons), and
$\omega_{\max}$ is the frequency cutoff.  The kernels are defined by
\begin{align}
K^F(\tau,\omega)
&=\frac{e^{-\tau\omega}}{1+e^{-\beta\omega}},
\label{eq:KF}\\
K^B(\tau,\omega)
&=\omega\,\frac{e^{-\tau\omega}}{1-e^{-\beta\omega}} .
\label{eq:KB}
\end{align}
By introducing dimensionless variables
$x=2\tau/\beta-1$, $y=\omega/\omega_{\max}$, and
$\Lambda=\beta\omega_{\max}$, they become
\begin{align}
k^F_\Lambda(x,y)&=\frac{e^{-\Lambda xy/2}}{2\cosh(\Lambda y/2)} ,
\label{eq:kF-dim}\\
k^B_\Lambda(x,y)&=y\,\frac{e^{-\Lambda xy/2}}{2\sinh(\Lambda y/2)} .
\label{eq:kB-dim}
\end{align}
We first analyze the fermionic kernel, returning to the bosonic case
at the end.  The fermionic IR basis is the SVD of
Eq.~\eqref{eq:kF-dim},
\begin{equation}
k^F_\Lambda(x,y)=\sum_{l=0}^{\infty} s_l(\Lambda)\,u_l(x)\,v_l(y) ,
\label{eq:SVD}
\end{equation}
and we focus on the effective rank of the singular values, which is
defined for a threshold $0<\epsilon<1$ as
\begin{equation}
N_{\rm eff}(\Lambda;\epsilon)
=\#\{\,l\,|\,s_l/s_0>\epsilon\,\} .
\label{eq:Neff}
\end{equation}

\textit{Reduction to a finite-Laplace-transform problem.}---We
regard Eq.~\eqref{eq:SVD} as the singular-value problem for the
integral operator
\begin{equation}
(\mathcal K^F f)(x)=\int_{-1}^{1} k^F_\Lambda(x,y)f(y)\,dy .
\label{eq:KF-operator}
\end{equation}
The left and right singular functions satisfy
\begin{equation}
\mathcal K^F v_l=s_l u_l,\qquad
(\mathcal K^F)^*u_l=s_l v_l .
\label{eq:svd-operator}
\end{equation}
For nonzero singular values, this is equivalent to the
self-adjoint eigenvalue problem
\begin{equation}
\mathcal K^F(\mathcal K^F)^*u_l=s_l^2u_l .
\label{eq:left-eigen}
\end{equation}
Once $u_l$ is obtained, the right singular function is
recovered as
\begin{equation}
v_l=s_l^{-1}(\mathcal K^F)^*u_l .
\label{eq:right-recovery}
\end{equation}

We introduce the rescaled, weighted Laplace
transform $B_\Lambda:L^2(-1,1)\to L^2(-\Lambda/2,\Lambda/2)$ and its
adjoint by
\begin{align}
(B_\Lambda \varphi)(u)
&=\frac{1}{\cosh u}\!\int_{-1}^{1}\!\!\! e^{-ut}\varphi(t)\,dt ,
\nonumber\\
(B_\Lambda^* g)(t)
&=\!\int_{-\Lambda/2}^{\Lambda/2}\!\!\frac{e^{-ut}}{\cosh u}\,g(u)\,du ,
\label{eq:B}
\end{align}
with $u\in[-\Lambda/2,\Lambda/2]$ and $t\in[-1,1]$.  Since
$\mathcal K^F(\mathcal K^F)^*=(2\Lambda)^{-1}
B_\Lambda^*B_\Lambda$, the eigenproblem for $u_l$ becomes
\begin{equation}
B_\Lambda^*B_\Lambda u_l=(\Lambda\lambda_l/2)u_l,
\qquad \lambda_l:=4s_l^2 .
\label{eq:B-eigen}
\end{equation}
The normalization identity used here is given in Appendix A.
With $g_l=B_\Lambda u_l$, the function $g_l$ is an
eigenfunction of $B_\Lambda B_\Lambda^*$ on the rescaled frequency
side with the same eigenvalue,
i.e.,
$B_\Lambda B_\Lambda^*g_l=(\Lambda\lambda_l/2)g_l$.

The thermal weight $1/\cosh u$ is buried inside the $u$-integral of
the kernel of $B_\Lambda^*B_\Lambda$
and cannot be removed by a simple change of variables on the
$t$ side (the rescaled imaginary-time side).
We therefore consider the isospectral operator
$B_\Lambda B_\Lambda^*$ on the rescaled-frequency side.  It is defined by
\begin{equation}
\begin{aligned}
(B_\Lambda B_\Lambda^*g)(v)
&=\!\int_{-\Lambda/2}^{\Lambda/2}\!\!
J^F_\Lambda(v,v')g(v')\,dv',\\
J^F_\Lambda(v,v')
&=\frac{1}{\cosh v\cosh v'}
\int_{-1}^{1}\!\! e^{-t(v+v')}dt .
\end{aligned}
\end{equation}
By writing the corresponding eigenfunction as
$g_l(v)=\cosh v\,H_l(v)$, we move the thermal weight from the
kernel to the right-hand side, where it appears as $\cosh^2 v$.
The result is the generalized eigenproblem
\begin{equation}
\int_{-\Lambda/2}^{\Lambda/2}\!\!\widetilde J_\Lambda(v,v')\,H_l(v')\,dv'
=\frac{\Lambda\lambda_l}{2}\cosh^2 v\,H_l(v) ,
\label{eq:gen-eigen}
\end{equation}
\begin{equation}
\widetilde J_\Lambda(v,v')
:=\bigl[\widetilde B_\Lambda\widetilde B_\Lambda^*\bigr](v,v')
=\frac{2\sinh(v+v')}{v+v'}.
\label{eq:Jtilde}
\end{equation}
Here
$(\widetilde B_\Lambda a)(v):=\int_{-1}^{1}e^{-tv}a(t)\,dt$ is the
\emph{weight-free} finite Laplace transform.  Thus the IR spectral
problem reduces to a generalized eigenproblem generated by
$\widetilde B_\Lambda$, with the kernel $\widetilde J_\Lambda$ on the
left-hand side and the weight $\cosh^2 v$ on the
right-hand side.

\begin{figure}[t]
\centering
\includegraphics[width=\columnwidth]{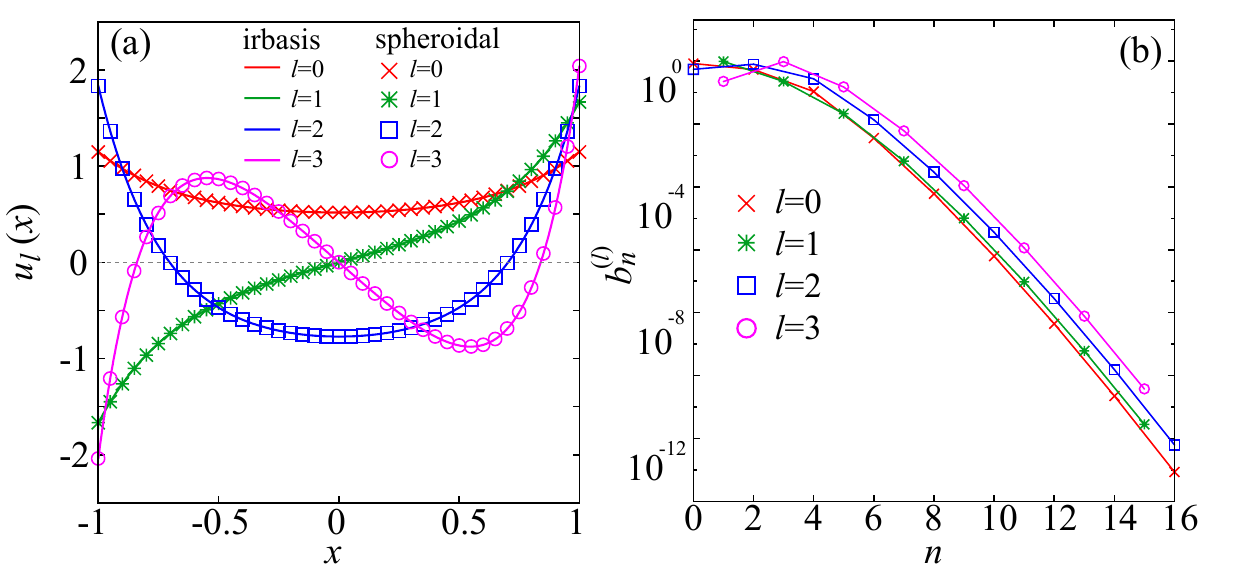}
\caption{Verification of the spheroidal representation in
Eq.~\eqref{eq:phi-spheroidal} at $\Lambda=10$.
(a) IR left singular functions $u_l(x)$ for $l=0,1,2,3$.
The solid lines are obtained from the SVD of the IR kernel
$k^F_\Lambda$ [Eq.~\eqref{eq:kF-dim}] using \texttt{irbasis}~\cite{Chikano2019}.  The symbols show the spheroidal reconstruction from
Eq.~\eqref{eq:phi-spheroidal} with truncation $n\le 16$.
The spheroidal reconstruction agrees well with the direct SVD results for all $l$.
(b) Magnitude of the spheroidal expansion coefficients
$b_n^{(l)}$ defined by
$u_l(x)=\sum_n b_n^{(l)}S_n(x;\Lambda)$ for normalized
$u_l$.  The coefficients decay rapidly with $n$.}
\label{fig:structure}
\end{figure}

\textit{Spheroidal representation of IR eigenfunctions.}---The two
operators $\widetilde B_\Lambda\widetilde B_\Lambda^*$
(acting on functions of $v$) and
$\widetilde B_\Lambda^*\widetilde B_\Lambda$
(acting on functions of $t$ in $L^2(-1,1)$) share the same
nonzero eigenvalues $\mu_n$.
The corresponding $t$-variable kernel is
\begin{equation}
\widetilde K_\Lambda(x,t)
=\int_{-\Lambda/2}^{\Lambda/2}\!\! e^{-v(x+t)}\,dv
=\frac{2\sinh[\Lambda(x+t)/2]}{x+t},
\label{eq:Ktilde}
\end{equation}
which has the same $\sinh z/z$ form as $\widetilde J_\Lambda$ but acts
on a different interval.  The operator $\widetilde B_\Lambda^*\widetilde B_\Lambda$ on the $t$ variable is more convenient for the spheroidal
analysis below because the second-order Sturm--Liouville operator
\begin{equation}
D_t=\partial_t\bigl((1-t^2)\partial_t\bigr)+(\Lambda/2)^2 t^2 ,
\label{eq:Dt}
\end{equation}
commutes with $\widetilde B_\Lambda^*
\widetilde B_\Lambda$ on $L^2(-1,1)$, as shown in
Appendix B.
Since this Sturm--Liouville operator has a simple spectrum, the
eigenfunctions of $\widetilde B_\Lambda^*\widetilde B_\Lambda$ can be
chosen to be eigenfunctions of $D_t$.
Its
eigenfunctions $S_n(t;\Lambda)$ are the oblate spheroidal
wave functions satisfying
$\widetilde B_\Lambda^*\widetilde B_\Lambda S_n=\mu_n S_n$.
Details of the oblate spheroidal wave functions and their Legendre expansion are
shown in Appendix C.
In the infinite-temperature
limit $\Lambda\to 0$, the $(\Lambda/2)^2 t^2$ term vanishes and
$D_t$ reduces to the Legendre operator, recovering the known
reduction of the IR basis to Legendre polynomials.

We expand $H_l=\sum_n c_n^{(l)}\Phi_n$ in Eq.~\eqref{eq:gen-eigen}
in terms of the orthonormal left singular functions
$\Phi_n=\mu_n^{-1/2}\widetilde B_\Lambda S_n$ in the $v$ variable.
This basis diagonalizes the weight-free core, so that the remaining
nondiagonal part comes only from the thermal weight.
Projecting Eq.~\eqref{eq:gen-eigen} onto this basis gives the matrix
eigenproblem
\begin{equation}
\begin{aligned}
\mu_n c_n^{(l)}&=\frac{\Lambda\lambda_l}{2}\sum_m W_{nm}c_m^{(l)},\\
W_{nm}&=\!\!\int_{-\Lambda/2}^{\Lambda/2}\!\!\Phi_n\cosh^2\!v\,\Phi_m\,dv ,
\end{aligned}
\label{eq:Wmatrix}
\end{equation}
which simultaneously determines the IR eigenvalues $\lambda_l=4 s_l^2$
and the expansion coefficients $c_n^{(l)}$.  Since
$\Phi_n(-v)=(-1)^n\Phi_n(v)$ and $\cosh^2 v$ is even, the matrix
problem separates into even and odd parity sectors.  The label $l$
denotes the eigenvalues after merging the two sectors in decreasing
order.  With this convention, $c_n^{(l)}=0$ unless $n$ and $l$ have
the same parity.
The eigenvalue identity
$\int_{-1}^{1}\!\widetilde K_\Lambda(x,t)S_n(t)dt=\mu_n S_n(x)$
together with $u_l(x)=(2/\Lambda\lambda_l)(\widetilde B_\Lambda^*
H_l)(x)$ gives the IR left singular functions as
\begin{equation}
u_l(x)=\frac{2}{\Lambda\lambda_l}\sum_{n=0}^{\infty}
c_n^{(l)}\sqrt{\mu_n(\Lambda)}\,S_n(x;\Lambda) ,
\label{eq:phi-spheroidal}
\end{equation}
This result shows that \emph{the IR eigenfunctions have a
weighted-spheroidal representation}.  The point is not merely that the
eigenfunctions are expanded in a complete basis, but that the oblate
spheroidal basis diagonalizes the weight-free finite-Laplace core, while
the remaining thermal weight fixes the coefficients through the
generalized matrix eigenproblem in Eq.~\eqref{eq:Wmatrix}.
The projection from Eq.~\eqref{eq:gen-eigen} to
Eq.~\eqref{eq:Wmatrix} and the reconstruction formula
Eq.~\eqref{eq:phi-spheroidal} are detailed in Appendix D.
In Fig.~\ref{fig:structure}, we compare the direct SVD with the
spheroidal reconstruction at $\Lambda=10$ and show the expansion
coefficients.  The good agreement confirms the spheroidal
reconstruction, and the rapid decay of the coefficients supports
the convergence of the expansion.

The above analysis establishes two key results.  First, the IR
eigenfunctions are given by an explicit series representation
[Eq.~\eqref{eq:phi-spheroidal}] in the oblate
spheroidal wave functions.  Second, the IR spectral problem reduces
to a generalized eigenproblem [Eq.~\eqref{eq:gen-eigen}] generated
by the \emph{weight-free finite Laplace transform} $\widetilde
B_\Lambda$.  The latter fact reveals the analytic origin of the
two empirical hallmarks of IR-kernel compression, namely the rapid
decay of $s_l$ with $l$ and the $\log\Lambda$ scaling of
$N_{\rm eff}$.  In what follows, we derive them by taking the
low-temperature limit of the generalized eigenproblem.

\textit{Low-temperature limit and Hilbert kernel.}---For the
$\Lambda\to\infty$ analysis of Eq.~\eqref{eq:gen-eigen},
it is enough to analyze the $v>0$ side for the leading
low-temperature rank count.  The $v<0$ side gives the same
Hilbert-kernel limit after $v\to -v$.
The two signs therefore give identical leading contributions to the
rank.  In the $v>0$
sector, we rescale
\begin{equation}
y=2v/\Lambda=\omega/\omega_{\max}\in(0,1]
\label{eq:y-def}
\end{equation}
and renormalize
\begin{equation}
H_l(v)=e^{-\Lambda y/2}\tilde h_l(y) .
\label{eq:H-renorm}
\end{equation}
In this scaling limit $\widetilde
J_\Lambda(v,v')=2\sinh(v+v')/(v+v') \sim
e^{\Lambda(y+y')/2}/(y+y')$ for $v+v'\gg 1$
($y+y'\gg\Lambda^{-1}$), while
$\cosh^2 v\sim e^{\Lambda y}/4$ on the right-hand side of
Eq.~\eqref{eq:gen-eigen}.  The diverging exponent $e^{\Lambda y/2}$ on
each side cancels after the rescaling in Eq.~\eqref{eq:H-renorm}.  Thus, in the
low-temperature limit, Eq.~\eqref{eq:gen-eigen} reduces to the
Hilbert-kernel equation
\begin{equation}
\begin{aligned}
\int_{c/\Lambda}^{1}\frac{\tilde h_l(y')}{y+y'}\,dy'
&=\eta_l\,\tilde h_l(y),\quad y\in[c/\Lambda,1],\\
\eta_l&=\frac{\Lambda\lambda_l}{8}
=\frac{\Lambda s_l^2}{2},
\end{aligned}
\label{eq:hilbert-IR}
\end{equation}
where $c=O(1)$ is a cutoff constant.

\begin{figure}[t]
\centering
\includegraphics[width=\columnwidth]{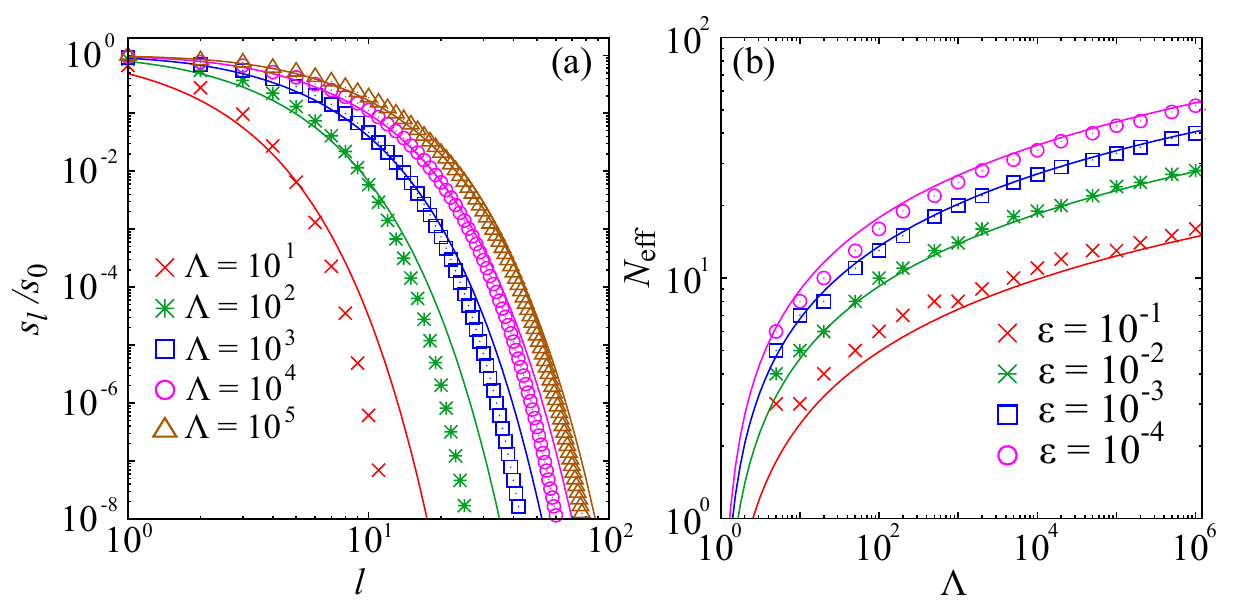}
\caption{Low-temperature scaling of IR singular values and effective rank.
(a) Normalized singular values $s_l/s_0$ as a function of $l$ for
several values of $\Lambda$.
Symbols are the direct
SVD of the kernel (\texttt{irbasis}~\cite{Chikano2019}).  Solid curves show the
asymptotic behavior
$1/\!\sqrt{\cosh(\pi k_l)}$ [Eq.~\eqref{eq:relative-sl}] with
$k_l=l\pi/(2\log\Lambda)$.
(b) Effective IR rank $N_{\rm eff}^{F}(\Lambda;\epsilon)$
[Eq.~\eqref{eq:Neff}] versus $\Lambda$ for
several values of $\epsilon$.
Symbols show the results obtained by the direct SVD and
solid lines show the analytic
prediction Eq.~\eqref{eq:rank-final},
$N_{\rm eff}^{F}\sim(2/\pi^2)\,\arcosh(\epsilon^{-2})\log\Lambda$.}
\label{fig:lowT}
\end{figure}

We now use the logarithmic variable $u=-\log y$ and write the
transformed function as
$g_l(u)=e^{-u/2}\tilde h_l(e^{-u})$.  The interval length is
$L=\log\Lambda+O(1)$, and
Eq.~\eqref{eq:hilbert-IR} becomes, up to endpoint errors that do not
change the leading count, the finite convolution problem
\begin{equation}
\int_0^L
\frac{g_l(u')}{2\cosh((u-u')/2)}\,du'
=\eta_l g_l(u),\quad u\in[0,L].
\label{eq:log-box}
\end{equation}

To take the bulk (low-temperature) limit, choose an interior point $u_0$ and write
$\xi=u-u_0$.  Then $u'\in[0,L]$ becomes
$\xi'\in[-u_0,L-u_0]$, which tends to $(-\infty,\infty)$ as
$L\to\infty$ with both $u_0$ and $L-u_0$ tending to infinity.
In the bulk infinite-line limit, the convolution operator
associated with Eq.~\eqref{eq:log-box} is diagonalized by plane waves
as
\begin{equation}
\int_{-\infty}^{\infty}
\frac{e^{-ik\xi'}}{2\cosh((\xi-\xi')/2)}\,d\xi'
=\frac{\pi}{\cosh(\pi k)}e^{-ik\xi}.
\label{eq:hilbert-cont-eigenvalue}
\end{equation}
For finite $L$, the continuous Fourier wavenumber in
Eq.~\eqref{eq:hilbert-cont-eigenvalue} is discretized by the finite
log-energy box.  A single sign sector has the effective logarithmic
length $\log\Lambda+O(1)$.  By considering both sign sectors, this
effective logarithmic length becomes
\begin{equation}
L_{\rm eff}=2\log\Lambda+O(1).
\label{eq:L-eff}
\end{equation}
From this, we obtain the leading asymptotic eigenvalue sequence as
\begin{equation}
\begin{aligned}
&\eta_l=\frac{\Lambda s_l^2}{2}
\approx\frac{\pi}{\cosh(\pi k_l)},\\
&
k_l=\frac{l\,\pi}{L_{\rm eff}}
\simeq\frac{l\,\pi}{2\log\Lambda}.
\end{aligned}
\label{eq:sl-from-eigenvalue}
\end{equation}

\textit{Asymptotic singular-value decay and effective rank.}---By
normalizing Eq.~\eqref{eq:sl-from-eigenvalue} by the largest singular
value, we obtain
\begin{equation}
\frac{s_l}{s_0}\approx
\frac{1}{\sqrt{\cosh(\pi k_l)}}.
\label{eq:relative-sl}
\end{equation}
For large $l$, this gives the asymptotic behavior
\begin{equation}
\frac{s_l}{s_0}\;\sim\;
\exp\!\left(-\frac{\pi^2\,l}{4\,\log\Lambda}\right).
\label{eq:sl-decay}
\end{equation}
Thus, the relative IR singular values decay exponentially in
$l$ with rate $\pi^2/(4\log\Lambda)$.
In Fig.~\ref{fig:lowT}(a), we compare Eq.~\eqref{eq:relative-sl} with
the direct SVD results.  The good agreement at large $\Lambda$ supports
the asymptotic analysis.

We next derive the effective rank from these asymptotic singular
values.  Imposing $s_l/s_0>\epsilon$ gives
$\cosh(\pi k_l)<\epsilon^{-2}$.  Using $k_l=l\pi/L_{\rm eff}$ and
$L_{\rm eff}=2\log\Lambda+O(1)$, we obtain
\begin{equation}
N_{\rm eff}^{F}(\Lambda;\epsilon)
\sim\frac{2}{\pi^2}\,\arcosh(\epsilon^{-2})\,\log\Lambda,
\label{eq:rank-final}
\end{equation}
which is the analytic origin of the empirical $\log\Lambda$
scaling of the IR rank.  We also compare Eq.~\eqref{eq:rank-final}
with the direct SVD results in Fig.~\ref{fig:lowT}(b) and confirm the
good agreement, which also supports our analysis.  We note that the
resulting effective rank can also be derived rigorously by applying
Widom's asymptotic analysis of the $1/\!\cosh$-type convolution kernel
on a finite interval~\cite{Widom1966Hankel} to Eq.~\eqref{eq:log-box}.

\textit{Bosonic case.}---In the bosonic case, changing only
the thermal weight in Eq.~\eqref{eq:gen-eigen} from $\cosh^2 v$ to
$(\sinh v/v)^2$ allows the same analysis.  The extra factor $y$
in $k^B_\Lambda$ introduces an additional low-energy weight.
With the same change of variables and the gauge
$H_{B,l}(v)=e^{-\Lambda y/2}y\,\tilde h_{B,l}(y)$, Eq.~\eqref{eq:gen-eigen}
reduces in the low-temperature limit to
\begin{equation}
\int_0^1\frac{y\,y'}{y+y'}\,\tilde h_{B,l}(y')\,dy'
=\eta_l^B(\infty)\,\tilde h_{B,l}(y),\quad y\in(0,1].
\label{eq:bose-hilbert}
\end{equation}
Here $\eta_l^B(\infty)$ denotes the eigenvalue of this limiting bosonic
operator.
The finite-$\Lambda$ lower cutoff $c/\Lambda$ can be sent to zero
because the kernel $yy'/(y+y')$ is integrable at $y=0$.  The
limiting kernel is square integrable, so the corresponding symmetric
integral operator is Hilbert--Schmidt, whose eigenvalues form a discrete
sequence accumulating only at zero.
After normalizing by the largest bosonic singular value,
the trivial $\Lambda$ dependence in $s_l^B(\Lambda)$ cancels.  Therefore, the
normalized singular values approach the following $\Lambda$-independent
limiting sequence,
\begin{equation}
\frac{s_l^B(\Lambda)}{s_0^B(\Lambda)}
\longrightarrow
\left(\frac{\eta_l^B(\infty)}{\eta_0^B(\infty)}\right)^{1/2}.
\label{eq:bose-normalized-limit}
\end{equation}
Since the eigenvalues of the limiting Hilbert--Schmidt operator
accumulate only at zero, this $\Lambda$-independent limiting sequence
contains only a finite number of terms above a fixed relative threshold.
Therefore, Eq.~\eqref{eq:bose-normalized-limit} yields the
saturation of the effective rank,
\begin{equation}
N_{\rm eff}^{B}(\Lambda;\epsilon)\longrightarrow
N_{\rm eff}^{B}(\infty;\epsilon)=O(1).
\label{eq:bose-rank-saturation}
\end{equation}
The contrast with the fermionic
case comes from the extra factor $y$ in the bosonic kernel, which
removes the logarithmic energy interval.

\textit{Summary and discussion.}---In this Letter, we have shown that
the IR singular-value problem reduces to a generalized eigenproblem
generated by a finite Laplace transform.  After the thermal weight is
moved out of the kernel, the associated integral operator has a
commuting second-order differential operator, leading to a series
representation of the IR singular functions in the oblate spheroidal
wave functions.
This finite-Laplace reduction clarifies that the fermionic and bosonic
cases share the same finite-Laplace backbone and the resulting oblate
spheroidal structure.  They differ only in the thermal
weight, $\cosh^2 v$ for fermions and $(\sinh v/v)^2$ for bosons.  In
the fermionic low-temperature limit, the reduced eigenproblem
near zero energy becomes a finite log-energy convolution problem with
the Hilbert kernel, yielding the
relative singular-value decay Eq.~\eqref{eq:sl-decay} and the rank law
Eq.~\eqref{eq:rank-final}.  For the bosonic case, the additional
factor of frequency induces the rank saturation
Eq.~\eqref{eq:bose-rank-saturation}.

The finite-Laplace transform at the core of the IR kernel is
closely related to the integral kernel in the time-bandwidth
concentration problem of Slepian, Pollak, and Landau~\cite{Slepian1961,LandauPollak1961,Slepian1983}.  The corresponding finite Fourier problem is governed by the sinc
kernel $\sin[c(t-t')]/[\pi(t-t')]$.  The IR finite-Laplace kernel is
obtained from it by the Wick rotation
$c\to i\Lambda/2$ together with the reflection $t'\to -t'$.
\begin{equation}
\frac{\sin[c(t-t')]}{\pi(t-t')}
\longrightarrow
\frac{\sinh[\Lambda(t+t')/2]}{\pi(t+t')}.
\label{eq:wick-sinc}
\end{equation}
The right-hand side is proportional to the
kernel of $\widetilde B_\Lambda^*\widetilde B_\Lambda$.  At the level
of the commuting differential operator, the same continuation changes
the prolate term $-c^2t^2$ into the oblate term
$(\Lambda/2)^2t^2$ in Eq.~\eqref{eq:Dt}.
In this sense, the IR kernel contains a hidden Wick-rotated
version of the Slepian structure, in which a natural integral operator
for information compression is connected to a classical second-order
differential equation from mathematical physics.
This Wick-rotated Slepian structure is the key reason why the IR spectral
problem admits a systematic analytic treatment.  The unexpected
correspondence opens a new route to analyzing the mathematical
structure underlying information compression in quantum many-body
problems.

\bigskip

\textit{Acknowledgments.}---The author thanks Hiroshi Shinaoka for
helpful discussions on the IR basis at an early stage of this study.
The author used AI-based tools,
including GPT-based tools (OpenAI) and Claude (Anthropic), as
interactive aids for exploring algebraic rearrangements.  These tools
helped suggest candidate transformations, including the factorization
route that led to the spheroidal representation.  All derivations,
calculations, and conclusions were independently verified by the
author, who takes full responsibility for the content.  This work was
supported by JSPS KAKENHI Grant No.~26K00652.  TM was supported by
JST FOREST JPMJFR236N.

\clearpage

\section*{End Matter}

\renewcommand{\theequation}{A\arabic{equation}}
\renewcommand{\theHequation}{A\arabic{equation}}
\setcounter{equation}{0}
\subsection*{Appendix A: Finite-Laplace Reduction}
\phantomsection\label{app:finite-laplace}
We explain the adjoint convention and the normalization used in
Eq.~\eqref{eq:B}.  For the fermionic kernel Eq.~\eqref{eq:kF-dim}, the
adjoint of $\mathcal K^F$ with respect to the standard $L^2(-1,1)$
inner product is given by
\begin{equation}
((\mathcal K^F)^*f)(y)=\int_{-1}^{1} k^F_\Lambda(t,y)f(t)\,dt .
\label{eq:KF-adjoint}
\end{equation}
Therefore
\begin{equation}
\begin{aligned}
(\mathcal K^F(\mathcal K^F)^*f)(x)
&=\int_{-1}^{1}\!\!dy\int_{-1}^{1}\!\!dt\,
\frac{e^{-\Lambda y(x+t)/2}}{4\cosh^2(\Lambda y/2)}
f(t)\\
&=\frac{1}{2\Lambda}\int_{-1}^{1}\!\!dt
\int_{-\Lambda/2}^{\Lambda/2}\!\!du\,
\frac{e^{-u(x+t)}}{\cosh^2u}f(t),
\end{aligned}
\end{equation}
where $u=\Lambda y/2$.  Hence
$\mathcal K^F(\mathcal K^F)^*=(2\Lambda)^{-1}B_\Lambda^*B_\Lambda$,
which is the normalization identity used in the reduction.

\renewcommand{\theequation}{B\arabic{equation}}
\renewcommand{\theHequation}{B\arabic{equation}}
\setcounter{equation}{0}
\subsection*{Appendix B: Commuting Differential Operator}
\phantomsection\label{app:commuting-differential}
We show that the Sturm--Liouville operator $D_t$ in Eq.~\eqref{eq:Dt}
commutes with $\widetilde B_\Lambda^*\widetilde B_\Lambda$.  Here
commutation means that, for smooth functions $f$,
\begin{equation}
D_x\!\left[(\widetilde B_\Lambda^*\widetilde B_\Lambda f)(x)\right]
=\bigl(\widetilde B_\Lambda^*\widetilde B_\Lambda D_t f\bigr)(x),
\label{eq:commutation-identity}
\end{equation}
where $D_x$ and $D_t$ denote the same differential expression
Eq.~\eqref{eq:Dt} acting on $x$ and $t$, respectively.  This
operator acts as
\begin{equation}
\begin{aligned}
&(\widetilde B_\Lambda^*\widetilde B_\Lambda f)(x)
=\int_{-1}^{1}\widetilde K_\Lambda(x,t)f(t)\,dt,\\
&\widetilde K_\Lambda(x,t)
=\int_{-\Lambda/2}^{\Lambda/2}e^{-v(x+t)}dv
=F(x+t),\\
&F(z)=\frac{2\sinh(\Lambda z/2)}{z}.
\end{aligned}
\end{equation}
In the calculation below, $F$, $F'$, and $F''$ are evaluated at $z=x+t$.
Since $F$ satisfies
\begin{equation}
zF''(z)+2F'(z)-\left(\frac{\Lambda}{2}\right)^2zF(z)=0,
\end{equation}
one obtains
\begin{equation}
\begin{aligned}
(D_x-D_t)F
&=(t-x)\left\{zF''+2F'-\left(\frac{\Lambda}{2}\right)^2zF\right\}\\
&=0,
\end{aligned}
\end{equation}
Therefore, $D_x\widetilde K_\Lambda(x,t)=D_t\widetilde K_\Lambda(x,t)$.
For smooth test functions $f$,
{\setlength{\abovedisplayskip}{4pt}
\setlength{\belowdisplayskip}{4pt}
\setlength{\abovedisplayshortskip}{4pt}
\setlength{\belowdisplayshortskip}{4pt}
\begin{equation}
\begin{aligned}
D_x\!\int_{-1}^{1}\widetilde K_\Lambda(x,t)f(t)dt
&=\int_{-1}^{1}D_t\widetilde K_\Lambda(x,t)f(t)dt\\
&=\int_{-1}^{1}\widetilde K_\Lambda(x,t)D_t f(t)dt ,
\end{aligned}
\end{equation}%
}%
where the second equality follows by integration by parts twice
in $t$, with the boundary term
$(1-t^2)(\partial_t\widetilde K_\Lambda f-\widetilde K_\Lambda f')$
vanishing at $t=\pm1$.  This proves the commutation of $D_t$ with
$\widetilde B_\Lambda^*\widetilde B_\Lambda$.
We note that more general nonsymmetric finite-Laplace kernels were studied by
Bertero and Gr\"unbaum~\cite{Bertero1985}.

\renewcommand{\theequation}{C\arabic{equation}}
\renewcommand{\theHequation}{C\arabic{equation}}
\setcounter{equation}{0}
\subsection*{Appendix C: Spheroidal Equation and Legendre Expansion}
\phantomsection\label{app:spheroidal-functions}
The oblate spheroidal wave functions used in the main text are the normalized
eigenfunctions of the Sturm--Liouville operator $D_t$, with
the eigenvalues $\alpha_n(\Lambda)$,
\begin{equation}
\begin{aligned}
&D_t S_n(t;\Lambda)=\alpha_n(\Lambda)S_n(t;\Lambda),\\
&\langle S_n,S_m\rangle_{[-1,1]}=\int_{-1}^{1}S_n(t;\Lambda)S_m(t;\Lambda)\,dt=\delta_{nm}.
\end{aligned}
\label{eq:sph-eigen}
\end{equation}
Here $\langle f,g\rangle_I$ denotes the $L^2$ inner product on the
interval $I$.
Equivalently,
\begin{equation}
\frac{d}{dt}\!\left[(1-t^2)\frac{dS_n}{dt}\right]
\!+\!\left[\left(\frac{\Lambda}{2}\right)^2t^2-\alpha_n(\Lambda)\right]S_n=0 .
\label{eq:sph-diffeq}
\end{equation}
This is the oblate spheroidal equation with angular order $m=0$ in
the normalization used here.  The spheroidal functions can be
expanded in the Legendre polynomials as
\begin{equation}
\begin{aligned}
S_{2p}(t;\Lambda)
&=\sum_{q=0}^{\infty}d_{2p,2q}(\Lambda)P_{2q}(t),\\
S_{2p+1}(t;\Lambda)
&=\sum_{q=0}^{\infty}d_{2p+1,2q+1}(\Lambda)P_{2q+1}(t).
\end{aligned}
\label{eq:sph-legendre}
\end{equation}
Here $P_n$ denotes the Legendre polynomial and $p$ is a nonnegative integer.
Details of the expansion coefficients $d_{nq}$ in
Eq.~\eqref{eq:sph-legendre} are given in Ref.~\cite{Flammer1957}.
At $\Lambda=0$, Eq.~\eqref{eq:sph-diffeq} reduces to the Legendre
equation and $S_n$ reduces to the normalized Legendre polynomial.

\renewcommand{\theequation}{D\arabic{equation}}
\renewcommand{\theHequation}{D\arabic{equation}}
\setcounter{equation}{0}
\subsection*{Appendix D: Projection to the Spheroidal Matrix Problem}
\phantomsection\label{app:spheroidal-projection}
We give the projection steps that connect the generalized
eigenproblem Eq.~\eqref{eq:gen-eigen} to the matrix problem
Eq.~\eqref{eq:Wmatrix} and to the reconstruction formula
Eq.~\eqref{eq:phi-spheroidal}.  We use the spheroidal functions
$S_n(t;\Lambda)$ introduced in Appendix C.  They are normalized in
$L^2(-1,1)$ and diagonalize the operator $\widetilde B_\Lambda^*\widetilde B_\Lambda$ on the $t$ variable as

\begin{equation}
\widetilde B_\Lambda^*\widetilde B_\Lambda S_n=\mu_n S_n,
\qquad
\langle S_n,S_m\rangle_{[-1,1]}=\delta_{nm}.
\end{equation}
For each nonzero eigenvalue $\mu_n$, the associated singular
function on the $v$ interval is
$\Phi_n=\mu_n^{-1/2}\widetilde B_\Lambda S_n$.  The eigenvalue equation
for $S_n$ shows that the functions $\Phi_n$ are orthonormal.  The same
definition also gives
$\widetilde B_\Lambda\widetilde B_\Lambda^*\Phi_n=\mu_n\Phi_n$.
We expand the function $H_l$ in the $v$ variable as
$H_l=\sum_n c_n^{(l)}\Phi_n$ and project Eq.~\eqref{eq:gen-eigen}
onto $\Phi_n$.  Then we obtain
\begin{equation}
\mu_n c_n^{(l)}
=\frac{\Lambda\lambda_l}{2}\sum_m
\left[
\int_{-\Lambda/2}^{\Lambda/2}
\Phi_n(v)\cosh^2v\,\Phi_m(v)\,dv
\right]c_m^{(l)} .
\end{equation}
This is Eq.~\eqref{eq:Wmatrix}.

We detail how to recover the IR left singular function.
The eigenvalue equation $B_\Lambda^*B_\Lambda u_l=(\Lambda\lambda_l/2)u_l$,
together with $g_l=B_\Lambda u_l$, gives
\begin{equation}
u_l=\frac{2}{\Lambda\lambda_l}B_\Lambda^*g_l.
\end{equation}
Substituting $g_l(v)=\cosh v\,H_l(v)$ into the definition of
$B_\Lambda^*$ in Eq.~\eqref{eq:B}, we obtain
\begin{equation}
(B_\Lambda^*g_l)(t)
=\int_{-\Lambda/2}^{\Lambda/2}\frac{e^{-vt}}{\cosh v}\,\cosh v\,H_l(v)\,dv
=(\widetilde B_\Lambda^*H_l)(t).
\end{equation}
Therefore $u_l=(2/\Lambda\lambda_l)\widetilde B_\Lambda^*H_l$.
Expanding $H_l=\sum_n c_n^{(l)}\Phi_n$ and using
$\Phi_n=\mu_n^{-1/2}\widetilde B_\Lambda S_n$ together with
$\widetilde B_\Lambda^*\widetilde B_\Lambda S_n=\mu_n S_n$, we obtain
$\widetilde B_\Lambda^*\Phi_n=\sqrt{\mu_n}\,S_n$, which yields
\begin{equation}
u_l(x)=\frac{2}{\Lambda\lambda_l}\sum_{n=0}^{\infty}
c_n^{(l)}\sqrt{\mu_n(\Lambda)}\,S_n(x;\Lambda).
\end{equation}
This is Eq.~\eqref{eq:phi-spheroidal}.

\end{document}